\documentclass[twocolumn,showpacs,preprintnumbers,amsmath,amssymb,prb]{revtex4}

\usepackage{graphicx}
\usepackage{dcolumn}
\usepackage{bm}
\usepackage{amsmath}

\begin{document}


\title{Observation of excited states in a graphene double quantum dot}

\author{F. Molitor}
\email{fmolitor@phys.ethz.ch}
\author{H. Knowles}
\author{S. Dr\"oscher}
\author{U. Gasser}
\author{T. Choi}
\author{P. Roulleau}
\author{J. G\"uttinger}
\author{A. Jacobsen}
\author{C. Stampfer} \thanks{new address: \textit{JARA-FIT and II. Institute of Physics, RWTH Aachen, 52074 Aachen, Germany}}
\author{K. Ensslin}
\author{T. Ihn}
\affiliation{Solid State Physics Laboratory - ETH Zurich, Switzerland}

\date{\today}

\begin{abstract}

We study a graphene double quantum dot in different coupling regimes. Despite the strong capacitive coupling between the dots, the tunnel coupling is below the experimental resolution. We observe additional structures inside the finite-bias triangles, part of which can be attributed to electronic excited dot states, while others are probably due to modulations of the transmission of the tunnel barriers connecting the system to source and drain leads.

\end{abstract}

\maketitle


\section{Introduction}

Double quantum dot structures are promising candidates for the implementation of solid state spin qubits \cite{Loss98,Cerletti05}. Double dots have been realized in many different material systems, as for example in GaAs heterostructures \cite{VanderWiel02}, semiconductor nanowires \cite{Fasth05, Pfund06, Choi09} and carbon nanotubes \cite{Biercuk05}, and the control of individual electrons and spins has been achieved\cite{Fujisawa02, Ono02, Elzerman04, Petta05, Koppens06}. Graphene has been predicted to be particularly well-suited for spin-based quantum information processing, because spin-orbit interaction and hyperfine interaction are expected to be much weaker than in the material systems mentioned above, leading potentially to much longer spin coherence times \cite{Trauzettel07, Fischer09}.
Significant progress has been made recently in the fabrication and the understanding of graphene-based nanostructures, as for example constrictions and quantum dots \cite{Han07, Chen07, Todd_08, Molitor_prb09, Stampfer_prl09, Gallagher09, Liu_prb09, Han09, Stampfer08, Schnez09, Ponomarenko08}. Also graphene double quantum dots have been demonstrated recently \cite{Molitor_apl09,Moriyama09}. For example, Liu et al. \cite{Liu09} showed the presence of excited states in a double dot created in a top-gated graphene nanoribbon. In this work, we demonstrate the presence of excited states in a side-gated graphene double dot structure with a different geometry, formed by etching the islands out of a graphene flake. We study in detail the coupling between the two dots for different gate voltages. We show that an in-plane magnetic field changes the excited states spectrum.

\section{Experimental details}

\begin{figure}[t!]
\includegraphics[width=\columnwidth]{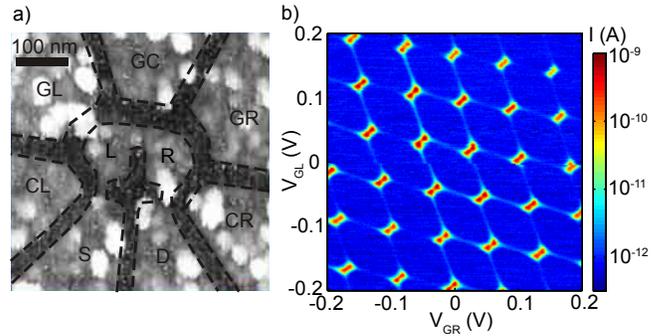}
\caption{(a) AFM image of the sample topology. The edges of the graphene regions are highlighted by dashed lines. The two dots, labelled by L and R, have a diameter of 90~nm and are seperated by a 30~nm wide constriction. The dots are connected by 20~nm wide constrictions to source and drain contacts (S and D). The global back gate and five in-plane graphene gates (CL, GL, GC, GR, CR) allow to tune the sample. (b) Current through the double dot measured as a function of $V_{\mathrm{GR}}$ and $V_{\mathrm{GL}}$ at $T=1.4$~K, $V_{\mathrm{BG}}=V_{\mathrm{GC}}=0~\mathrm{V}, V_{\mathrm{CR}}=-0.1~\mathrm{V},  V_{\mathrm{CL}}=-3.25~\mathrm{V}$ and  $V_{\mathrm{bias}}=-0.5~\mathrm{mV}$.}
\label{fig1}
\end{figure}

The sample consists of a double dot structure carved out of a graphene flake. Mechanical exfoliation of natural graphite flakes, followed by deposition onto a highly doped silicon wafer covered by 285 nm of silicon dioxide, is used to produce the graphene flakes. Thin flakes are identified with an optical microscope, and Raman spectroscopy is used to make sure that the flake consists only of one single graphene layer \cite{Ferrari06, Graf07}. The selected flake is contacted using electron beam lithography (EBL) and metal evaporation (Cr/Au). Finally it is patterned into the double dot structure shown in Fig.~\ref{fig1}(a) by a second EBL step and reactive ion etching based on argon and oxygen. 

The transport experiments were carried out in a variable temperature insert at 1.4 K, and at 120 mK base temperature of a standard $^{3}$He/$^{4}$He dilution refrigerator. In total, measurements of three different cool-downs are presented. Even if some details changed from one cool-down to the other, the main features presented in this work were present in every cool-down.

\section{Results and discussion}

Fig. \ref{fig1}(b) shows a measurement of the charge stability diagram, recorded at $T=1.4$~K. The hexagon pattern characteristic for double dots is clearly visible, and uniform over many double dot charge configurations. The current is maximal at the triple points, where the electrochemical potentials in both dots are aligned with each other and with the Fermi energy in the leads. These triple points are connected by faint lines of much smaller current, originating from inelastic cotunneling processes. Along these lines, the energy level in one dot is aligned with the electrochemical potential in the corresponding lead. Such well controlled double dot behavior, sometimes with less symmetric barriers, could be observed in the whole accessible range of positive back gate voltages (0-30~V) for all three cool-downs, as long as the barriers were not too closed to allow current detection.

\begin{figure}[t!]
\includegraphics[width=\columnwidth]{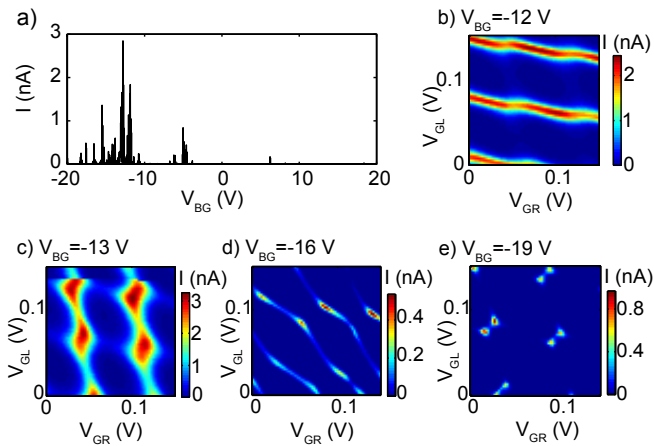}
\caption{$T$=1.4~K, $V_{\mathrm{bias}}=0.5~$mV$\approx4k_{\mathrm{B}}T$ (a) Current as a function of $V_{\mathrm{BG}}$, all the other gates at 0~V. (b)-(e): Current as a function of the voltage applied to GR and GL for different negative values of back gate voltage.}
\label{fig2}
\end{figure}

For negative back gate voltages, however, the situation can be quite different. This difference can be observed in Fig. \ref{fig2}(a), displaying the current through the double dot as a function of back gate voltage at 1.4~K. While the current is completely suppressed for positive values of $V_{\mathrm{BG}}$, resonances can be observed at negative gate voltages. Charge stability diagram measurements give a better understanding of this region. Fig. \ref{fig2}(b)-(e) represent a selection of such measurements for four different values of $V_{\mathrm{BG}}$ between -10~V and -20~V, where the resonances are strongest. They correspond to completely different situations: at $V_{\mathrm{BG}}=-12$~V, represented in Fig. \ref{fig2}(b), the current is high along the horizontal cotunneling lines and almost continuous across the triple points. Fig. \ref{fig2}(c) represents the opposite situation, occurring at $V_{\mathrm{BG}}=-13$~V, with high current at the triple points and along the vertical cotunneling lines. At $V_{\mathrm{BG}}=-16$~V [Fig. \ref{fig2}(d)], the regions of finite current describe diagonal, wavy lines, while at $V_{\mathrm{BG}}=-19$~V [Fig. \ref{fig2}(e)] a clean double dot charge stability diagram with current only at the triple points, is recovered. 

These different cases represent different coupling regimes between the two dots, and between the dots and the leads. In the case of $V_{\mathrm{BG}}=-12$~V, the current is high whenever the energy level in the left dot is aligned with the chemical potential in the left lead. This can be understood assuming the coupling between the right dot and the right lead is very strong compared to the coupling of the left dot to the leads and to the right dot, and therefore transport is dominated by the left dot. For $V_{\mathrm{BG}}=-13$~V, the opposite situation is realized, with strong coupling between the left dot and the left lead. Fig \ref{fig2}(d), recorded at $V_{\mathrm{BG}}=-16$~V, corresponds to a more symmetric situation, where the current along the cotunneling lines in both directions is almost equally strong. The coupling between both dots $E^{\mathrm{m}}_{\mathrm{C}}$ is very strong compared to the charging energies of the individual dots $E^{\mathrm{R}}_{\mathrm{C}}$ and $E^{\mathrm{L}}_{\mathrm{C}}$ ($E^{\mathrm{m}}_{\mathrm{C}}\approx0.5\cdot E^{\mathrm{R}}_{\mathrm{C}}\approx0.5\cdot E^{\mathrm{L}}_{\mathrm{C}}$), leading to almost diagonal lines, which would correspond to one large dot delocalized over both islands. Finally, at $V_{\mathrm{BG}}=-19$~V, the situation corresponds again to a well defined double dot, with all three tunnel barriers well closed ($E^{\mathrm{m}}_{\mathrm{C}}\approx0.2 E^{\mathrm{R}}_{\mathrm{C}}\approx0.2 E^{\mathrm{L}}_{\mathrm{C}}$).

\begin{figure}[t!]
\center
\includegraphics[width=0.9\columnwidth]{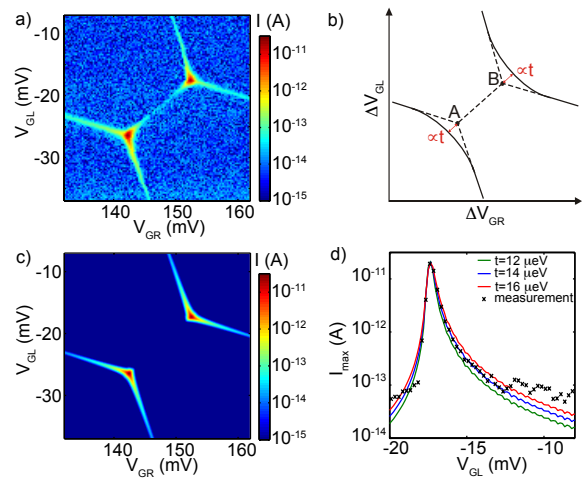}
\caption{(a) Measurement of a charge stability diagram around one pair of triple points at small bias voltage $V_{\mathrm{bias}}=15~\mu V$ and $V_{\mathrm{BG}}=-8~\mathrm{V}$, $V_{\mathrm{CL}}=V_{\mathrm{CR}}=0~\mathrm{V}$, $V_{\mathrm{CG}}=-0.5~\mathrm{V}$ and $T=120$~mK. (b) Schematic drawing of a charge stability diagram for two tunnel coupled quantum dots around two triple points labelled A and B. The corners of the hexagons, at the triple points, are rounded due to tunnel coupling, with a deviation from the straight, dotted lines proportional to the tunnel coupling strength t. (c) Simulation of a pair of triple points for $t=14~\mu$eV, and the energy independent part of the tunneling rates to the leads $\gamma_{\mathrm{L}}=1.26~$GHz and $\gamma_{\mathrm{R}}=1.69$~GHz. (d) Maximum current along the cotunneling lines in the range of the upper triple point for each value of $V_{\mathrm{GL}}$ for the measurement (black crosses) and simulations for $t=12~\mu$eV (green), $t=14~\mu$eV (blue) and $t=16~\mu$eV (red). }
\label{fig3}
\end{figure}

Fig. \ref{fig3}(a) shows a measurement for a charge stability diagram in the vicinity of one pair of triple points. It has been recorded at low temperature $T\approx120$~mK and at low bias voltage $V_{\mathrm{bias}}=15~\mu$V to prevent an expansion of the triple points to triangles. A negative back gate voltage $V_{\mathrm{BG}}=-8$~V has been chosen, because only in this regime the cotunneling lines are visible even at this low bias voltage. A corresponding schematic drawing of a charge stability diagram for two tunnel coupled quantum dots can be seen in Fig. \ref{fig3}(b). The tunnel coupling leads to rounded edges of the hexagons, with the point of charge balance shifted from the original triple point proportional to the strength of the tunnel coupling. Contrary to the situation depicted in Fig. \ref{fig3}(b), no rounding of the corners is visible in the measurement. This allows us to estimate an upper bound for the tunnel coupling $t\leq20~\mu$eV. This energy scale is comparable to the temperature broadening of the cotunneling lines ($k_{\mathrm{B}}T\approx10~\mu$eV) and about two orders of magnitude smaller than the capacitive coupling energy $E_C^{m}\approx1.3$~meV. The lever arms necessary for the determination of these energy scales were extracted from a measurement of the same pair of triple points at $V_{\mathrm{bias}}=1$~mV.
Fig. \ref{fig3}(c) shows the result of a numerical calculation of the current based on the rate equation using the lever arms and charging energies deduced from the measurement. Best agreement is found for $\gamma_{\mathrm{L}}=1.26~$GHz and $\gamma_{\mathrm{R}}=1.69$~GHz for the energy-independent part of the tunneling rates to the leads, and $t=14~\mu$eV for the tunnel coupling between both dots \cite{Gasser09, Gustavsson08}. A more detailed description of the calculation can be found in Ref. \cite{Gasser09}. Measurement and simulation are quite similar, except for the finite current measured along the line connecting both triple points, which is unexpected and can not be reproduced with this simple model. Fig. \ref{fig3}(d) shows a closer comparison between the measurement and the calculation. For each value of $V_{\mathrm{GL}}$ in the range of the upper triple point, the maximum current in the region of the cotunneling line is plotted for the measurement and for calculations with $t=12~\mu$eV, $t=14~\mu$eV and $t=16~\mu$eV. The tunnel coupling determines how fast the current drops as one goes away from the triple point along the cotunneling lines. Best agreement is found for $t=14~\mu$eV. However, due to the uncertainty in the lever arms in this regime and in the electronic temperature, the tunnel coupling strength can only be determined up to a factor of two. The difference between measurement and calculation in the tails of the peak far away from the triple point arises from the fact that the peak current from the measurement does not take values lower than the noise level.

Despite the strong capacitive coupling between the dots, the tunnel coupling is low. For the observation of Coulomb blockade a resistance of the order of $h/e^{2}$ is required. This resistance may arise from a tunnel barrier, as it is usually the case for GaAs-based quantum dot systems. In graphene, this tunnel coupling may be weak if there is a narrow but high barrier separating the dots. Such a situation could give rise to strong capacitive coupling (see Fig. \ref{fig2}(d)) while the tunnel coupling itself remains below the experimental resolution. Additional resonances in the central constriction \cite{Todd_08, Molitor_prb09, Stampfer_prl09, Gallagher09, Liu_prb09, Han09} and interactions might lead to an even more complicated situation. 

\begin{figure}[t!]
\includegraphics[width=\columnwidth]{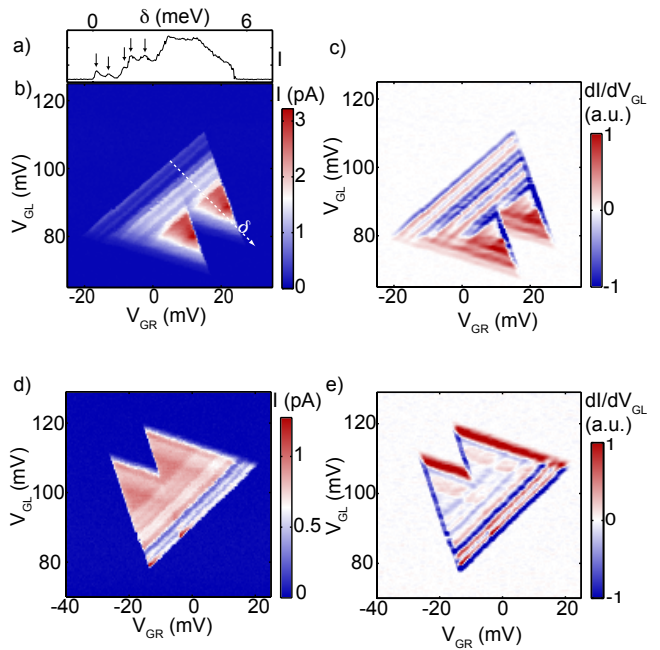}
\caption{Closer look at one pair of triple points at $T=120$~mK. (a) Current along the detuning axis. The arrows indicate the position of excited states. (b), (d): Current as a function of $V_{\mathrm{GR}}$ and $V_{\mathrm{GL}}$ for $V_{\mathrm{bias}}$=6~mV (b) and $V_{\mathrm{bias}}$=-6~mV (d). The dashed line in (b) represents the detuning line chosen for the measurement in (a). (c),(e): Corresponding representations of the current, numerically derivated by $V_{\mathrm{GL}}$ after smoothing over three data points.}
\label{fig4}
\end{figure}

Fig. \ref{fig4} displays a closer look at one pair of triple points for a finite applied bias voltage of $V_{\mathrm{bias}}=\pm6$~mV. This measurement is recorded at $V_{\mathrm{BG}}=25$~V, a region where the dot-lead coupling strengths are weak, and therefore no cotunneling lines are visible. Because of the high bias voltage, the triple points evolved into triangles \cite{VanderWiel02}. The extent of these triangles gives the lever arms needed to determine the energy scales of the double dot system.

Table \ref{tab} gives an overview of the main lever arms and energy scales. The symmetry of the structure is remarkable: the lever arms of both dots, as well as both single dot charging energies, are very similar. The charging energies are considerably higher than those reported in Ref. \cite{Liu09} (single dot charging energies $\approx3$~meV, mutual coupling energy $\approx0.4$~meV) despite the fact that these dots are slightly smaller. This is consistent with the fact that in the case of Ref. \cite{Liu09} the sample is partly covered by top gates, leading to increased screening.

\begin{table}
\caption{Overview of the main lever arms and energy scales. The lever arms and the mutual charging energy $E_{\mathrm{C}}^{\mathrm{m}}$ are determined from Fig. \ref{fig4}(b). The single dot charging energies are determined from Fig. \ref{fig1}(a) using the lever arms of this table.}
\label{tab}
\begin{center}
\begin{tabular}{ccccccc}
$\alpha_{\mathrm{GR,R}}$ & $\alpha_{\mathrm{GL,L}}$ & $\alpha_{\mathrm{GR,L}}$ & $\alpha_{\mathrm{GL,R}}$\\ 
$0.18$ & $0.20$ & $0.062$ & $0.066$\\
 &  &  & \\
$E_{\mathrm{C}}^{\mathrm{R}}$ & $E_{\mathrm{C}}^{\mathrm{L}}$ & $E_{\mathrm{C}}^{\mathrm{m}}$ & level spacing \\
 13.0~meV & 14.0~meV  & $2.4$~meV & $\approx0.5$~meV 
\end{tabular}
\end{center}
\end{table}

Inside the triangles of finite current, additional parallel lines can be seen. These lines are even clearer when plotting the derivative of the current along the $V_{\mathrm{GL}}$-axis, taken numerically after smoothening over 3 data points. The most prominent lines run parallel to the baseline of the triangles. Along such a line, the detuning between the energy levels in both dots is kept fixed. These lines are usually attributed to excited states in the right (left) dot for positive (negative) bias voltage. The lines can also be clearly seen in a cut along the detuning line (arrow in Fig. \ref{fig4}(a)). They have a typical level spacing of $\approx0.5$~meV for excited states in the right dot, and 0.4-0.8 meV in the left dot, and are much broader than $k_{\mathrm{B}}T\approx10~\mu$eV as a result of inelastic tunneling processes.

However, at closer inspection additional lines parallel to the lower edge of the triangle for $V_{\mathrm{bias}}=6$~mV and to the upper edge for $V_{\mathrm{bias}}=-6$~mV are visible. Along these lines, the alignment between the energy level of the left dot and the Fermi energy in the left lead is kept constant. In the case of negative bias voltage, these lines can not originate from an excited state in the left dot, assuming the number of carriers in both dots to stay constant. These lines are probably due to modulations of the tunneling coupling between the left dot and the left lead, because of resonances in this constriction \cite{Molitor_prb09}. These lines are broader than the lines parallel to the baseline, and only occur parallel to the nearly horizontal edge of the triangle, which corresponds to the direction of the stronger cotunneling lines. Parallel to the other edge of the triangles, no lines are observable, even when taking the derivative in the other direction.

\begin{figure}[t!]
\includegraphics[width=\columnwidth]{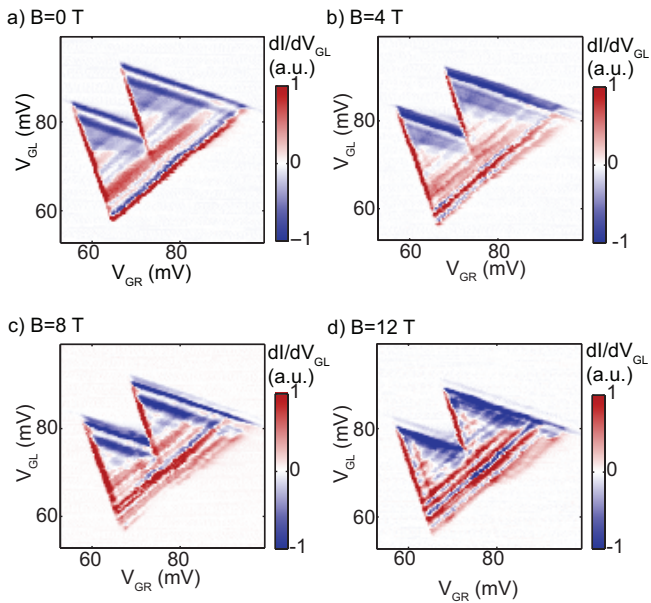}
\caption{$dI/dV_{\mathrm{GL}}$ as a function of $V_{\mathrm{GR}}$ and $V_{\mathrm{GL}}$ for $T=120$~mK and $V_{\mathrm{bias}}=-4$~mV for different values of in-plane magnetic field. $dI/dV_{\mathrm{GL}}$ is measured with a lock-in amplifier by adding an ac modulation of $200~\mu$V to $V_{\mathrm{GL}}$ and detecting the ac component of the current.}
\label{fig5}
\end{figure}

Fig. \ref{fig5} presents a study of one pair of triple points for different values of the magnetic field, oriented parallel to the graphene plane. The figure displays the current differentiated by $V_{\mathrm{GL}}$, measured directly by applying an ac modulation to $V_{\mathrm{GL}}$, and recording the ac current. Again, states parallel to the baseline as well as lines parallel to the upper edge of the triangle are visible. The position of the triangles in the gate voltage plane almost does not change at all up to $B=12$~T. This is in contrast to the case of a perpendicular magnetic field, where the position of the triple points and the intensities change significantly on a magnetic field scale $\Delta B\approx 250$~mT due to the effects of the field on the orbital part of the wavefunctions (not shown). Two effects of the parallel magnetic field on the triple points can be observed. First, with increasing magnetic field, the number of visible states parallel to the baseline increases, which is most pronounced for $B=12$~T [Fig. \ref{fig5}(d)]. This effect was observable for both pairs of triple points which were studied, and could originate from Zeeman splitting. However, it was not possible to analyze in detail the appearance of these additional lines because their broadening is similar to their spacing. The second effect is the appearance of a line parallel to the left edge of the triangle at high magnetic fields. This line originates from modulated transmission between the right dot and the right lead. The appearance of this line with high magnetic field is surprising, as one would not expect the in-plane magnetic field to localize states. We speculate that some areas of the structure are exposed to a finite component of the out-of plane field, owing to ripples always present in graphene flakes \cite{Lundeberg09}.

\section{Conclusion}

We have studied a graphene double quantum dot in different coupling regimes. Despite the strong capacitive coupling between both dots, the tunnel coupling is below the experimental resolution and no roundening of the hexagones at the triple points can be resolved. A numerical calculation of the current based on the rate equation leads to an estimation for the tunnel coupling of $t\approx14~\mu$eV. Inside the finite-bias triangles, additional structures can be observed, which we attribute to excited dot states, but partly also to imperfections in the tunnel barriers. With the application of an in-plane magnetic field, additional states become visible within the finite-bias triangles.


\begin{acknowledgments}
We thank B. K\"ung for helpful discussions, Y. Komijani for help with the setup and the Swiss National Foundation (SNF) and NCCR Nanoscience for financial support.
\end{acknowledgments}

\end{document}